# Avoiding the "Condorcet paradox" in stakeholder networks for participatory decision-making: an agent-based approach

*Michela Le Pira[1], Giuseppe Inturri[1], Matteo Ignaccolo[1],*
*Alessandro Pluchino[2], Andrea Rapisarda[2]*

[1] DICAR, Università di Catania, Italy
[2] Dipartimento di Fisica e Astronomia, Università di Catania, and INFN sezione di Catania, Italy

## Abstract

We address the problem of a participatory decision-making process where a shared priority list of alternatives has to be obtained while avoiding inconsistent decisions, related to the "Condorcet paradox". An agent-based model (ABM) is proposed to mimic this process in different social networks of stakeholders who interact according to an opinion dynamics model. Simulations' results show the efficacy of interaction in avoiding the paradox and finding a transitive and, above all, shared decision. These findings are in agreement with real participation experiences regarding transport planning decisions and can give useful suggestions on how to plan an effective participation process for sustainable policy-making based on opinion consensus.

## 1. INTRODUCTION

When more actors are involved in a decision-making process, individual preferences, in the form of an ordered list of alternatives, have to be aggregated into a unique collective decision. Among the many different aggregation methods that can be used to this aim (List, 2013), one of the most commonly adopted is the so called Pairwise Majority Rule (PMR). Nevertheless, the result of aggregation by the PMR can be an intransitive collective preference list, falling into the so called "Condorcet paradox" or "Condorcet cycle". It was studied for the first time in 1785 by the Marquis de Condorcet (1785) who demonstrated that, when more than two alternatives have to be ranked (for example A, B and C), the collective preference list can be intransitive (A>B>C>A), even if the individual preference orders are transitive (A>B>C). The final consequence is the impossibility of taking a consistent decision.

In this paper we present an agent-based model (ABM) able to mimic a participatory decision-making process where a given number of stakeholders, linked by a social network, exchange opinions (represented by their individual lists of preferences) in order to find a shared and transitive collective decision (i.e., a collective list). The model can be applied to all the decision-making contexts that involve participation in complex decisions with widespread collective impact, e.g. transport policy-making, environmental policy-making, land use planning, etc.

Lack of public participation in decision-making processes is considered one of the main causes of projects' failure (Marincioni and Appiotti, 2009), as confirmed by the wide diffusion of public infrastructure projects strongly opposed by the local community protests, with the so called

"NIMBY" syndrome ("Not In My BackYard"). It is now widely recognized the importance of involving all the actors (i.e. the stakeholders) along the decision-making processes in order to choose decisions which reflect the real needs and opinions of the interested parties (i.e. "most shared" solutions) rather than just the technically "best" ones. This is particularly true for policy-making at the urban scale, where citizens reflect a great variety of interests and opinions.

Public policy-making, in general, requires quantitative methods to evaluate to what degree each alternative satisfies a set of criteria selected to measure the achievement of objectives. Moreover, the decision-maker has to choose or make rankings among more than two alternatives (policies, projects or objectives). In such cases, decision-support methods (DSMs) such as Cost Benefit Analysis (CBA), or Multi Criteria Analysis (MCA) are considered adequate to assess the convenience of each alternative (Cascetta, 2009). Finally, the decision-maker has to prioritize the alternatives, usually by a ranking scale. When it comes to participatory decision-making processes, the problem of aggregating individual preferences is not trivial, being largely investigated by social choice theory (Arrow, 1951; Elster and Hylland, 1986). Indeed, the aggregation of individual stakeholders' preferences could determine an inconsistent decision and, most of all, it could not satisfactorily reflect stakeholders' expectations. In this respect, the aim of the model here presented is twofold: first, try to understand how it is possible to avoid, through the opinions interaction, the decision deadlock that can arise from the aggregation of the individual preferences with the PMR; second, try to find a collective transitive list able to satisfy stakeholders to a high degree, i.e. to find a highly shared collective decision.

The paper is organized as follows. In Section 2 the "Condorcet paradox" is introduced together with some strategies elaborated in order to avoid it. In section 3 an ABM is proposed to analyse some basic conditions to overcome the paradox, whilst assuring an acceptable degree of collective consensus. There we discuss also the simulations performed and the results together with some preliminary considerations about model validation. Finally, Section 4 provides some general conclusions.

## 2. THE PROBLEM OF THE "CONDORCET PARADOX" IN PARTICIPATORY DECISION-MAKING PROCESSES

Let us start with a simple example showing the appearance of the Condorcet paradox when the Pairwise Majority Rule is adopted for generating a collective preference list by aggregating individual ones. Consider five stakeholders ($N = 5$), each one with an individual list of four alternatives A, B, C and D ($n = 4$). Using PMR, the ranking of the collective list is obtained by computing how many times each alternative in a pair is preferred to the other one. In particular, the pairwise preferences of each individual list are coded as components of a binary vector assuming the values of +1 and -1[1]. Finally, the collective list is derived by applying a majority rule to the binary vectors.

As already anticipated, the aggregation of single transitive preference lists by the PMR does not exclude the possibility of intransitive collective lists as result, thus violating the ordering axiom (Arrow, 1951). Actually, by applying the PMR to the individual preference orders shown in Table 1, the final result will be what is called a "Condorcet cycle", i.e. A>B>C>D>A.

---

[1] As an example, for the couple of alternatives AB, if A is preferred to B then AB= +1, vice versa AB= -1.

**Table 1** Condorcet cycle resulted from aggregation of preference orders by PMR (sh = stakeholder).

| sh | Preference order | AB | AC | AD | BC | BD | CD |
|---|---|---|---|---|---|---|---|
| 1 | A>B>C>D | +1 | +1 | +1 | +1 | +1 | +1 |
| 2 | D>A>B>C | +1 | +1 | -1 | +1 | -1 | -1 |
| 3 | B>C>D>A | -1 | -1 | -1 | +1 | +1 | +1 |
| 4 | D>A>C>B | +1 | +1 | -1 | -1 | -1 | -1 |
| 5 | B>A>C>D | -1 | +1 | +1 | +1 | +1 | +1 |
|  | *PMR result:* | +1 | +1 | -1 | +1 | +1 | +1 |
|  | *Collective list:* | A>B>C>D>A |  |  |  |  |  |

It is easy to demonstrate that the probability of falling into a "Condorcet cycle" increases with the number of alternatives: given $n$ alternatives, then $n!$ possible transitive orders do exist, while $n*(n-1)/2$ is the number of pairs and, therefore, $2^{n*(n-1)/2}$ are all the possible - transitive and intransitive - binary vectors. Therefore, the probability to have a transitive list will be $P(n) = \frac{n!}{2^{n*(n-1)/2}}$ and it rapidly decreases with $n$, going to 0 when $n \to \infty$ (Fig. 1a). A helpful way to visualize the increasing asymmetry between the number of transitive and intransitive lists when $n$ increases is shown in Fig. 1b, representing a simplified pictorial view of the "collective preference space": the black cells in the grid indicate the $n!$ transitivity "islands" randomly distributed over the much larger intransitivity "sea", represented by the white cells (for $n = 6$, only $n! = 720$ transitive lists out of 32748 possible lists exist).

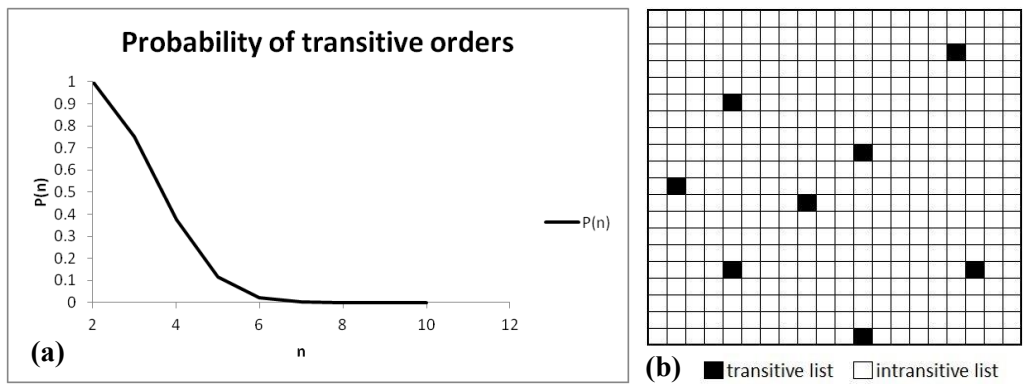

**Fig. 1.** (a) Probability P(n) of transitive orders as a function of the number of alternatives n. (b) Collective preference space: the few transitive lists are represented as black "islands" within the "sea" of intransitive lists.

It has been demonstrated that the occurrence of the paradox increases also with the number of agents, for example in "a large population of non interacting voters" (Raffaelli and Marsili, 2005). On the other hand, the result changes if agents interact before deciding. Actually, interaction is at the basis of most of the traditional participation tools (Rowe and Frewer, 2000), even in the form of the "remote" and anonymous interaction of the Delphi method (Dalkey and Helmer, 1963). In this respect, Raffaelli and Marsili (2005) demonstrated that, the larger the number of alternatives, the easier is to have a collective transitive list in an interacting population. Columbu et al. (2008) also show the effectiveness of interaction by evaluating the probability of collective transitive lists as a function of an interaction range and find out an optimal distance (namely a Kemeny distance) among agents able to reduce the probability of "Condorcet paradox". The widespread diffusion of digital social media allows today more and more people to easily interact and exchange opinions. On the other hand, the group decision-making based on decision-support methods usually provides averaged collective decisions which can be very far from each individual preference.

Summarizing, these models present two main drawbacks:
- they do not explicitly represent the real structure of the relationships among agents;
- they do not adequately consider the degree of consensus of the single agent against the final collective decision.

In order to overcome these limits and contextualize the problem in the field of collective decisions in sustainable policy-making, we present in the following an ABM which is able to reproduce the opinion dynamics interaction among stakeholders, linked by social networks with different topologies. We show this model allows both to circumvent the "Condorcet paradox" and find a shared transitive collective decision, i.e. a final collective list with a high similarity with the individual ones.

## 3. THE AGENT BASED MODEL

Opinion dynamics models have been largely used in the last years to capture the opinion evolution of groups of interacting agents, in order to investigate the underlying forces leading towards consensus formation, both on regular and complex network topologies (Hegselmann and Krause, 2002; Sznajd-Weron and Sznaid, 2000; Galam, 2002; Pluchino et al., 2005). A comprehensive review of the state of the art of social dynamics and opinion dynamics models based on statistical physics can be found in Castellano et al. (2009). Computer simulations play an important role in the study of social dynamics and one of the most successful methodologies used is Agent-based simulations. The latter are also widely used for applications to social and economic systems and they are also used as a tool to foster stakeholder engagement and collaboration (Voinov and Bousquet, 2010).

In a previous study (Le Pira et al., 2013), an ABM was already implemented to simulate the opinion dynamics on a particular stakeholder network where a binary decision has to be taken about a single transport policy measure: the various agents interact with each other and the conditions leading to the convergence of opinions according to a majority rule were investigated.

In this paper, a new ABM is presented to reproduce interaction in stakeholder networks with different topologies, where each stakeholder has an individual preference list over a set of (more than two) alternatives, and a collective preference list has to be found (see also Le Pira et al., 2015a). The aim is to see how a repeated interaction can help to escape from intransitive solutions and to foster a convergence of opinions towards a collective consistent solution with a high degree of consensus. The proposed ABM is here enriched by further exploring the parameter space and by considering other possible ways, different from interaction, to "escape" from the paradox: in this respect, we compare the results obtained through dynamical interaction among agents with other two different strategies, devoid of interaction, in order to see if they are equally successful in both avoiding the paradox and finding a transitive shared solution. Furthermore, the model will be used to reproduce real participation experiences, representing a first attempt of model validation.

Our model has been built within the software programming environment NetLogo (Wilensky, 1999), particularly suitable for agent-based modelling and simulations.

### 3.1 The ABM algorithm

Let's first describe the setup of the agent based model, common to all the strategies considered.

The $N$ stakeholders $S(i)$ (i=1,...,N) are represented as nodes of an undirected network, according to a selected topology with fixed connectivity. A set of alternatives is given and a preference list is randomly assigned to each stakeholder, thus representing his/her opinion. In addition, an integer random variable $I(i)$ is assigned to each stakeholder $S(i)$ to quantify his/her influence, i.e. the capability of influencing the opinions of his/her directly connected nodes (first neighbours) in the network. As explained in section 2, each preference list is transformed into a binary vector (see table 1), and the collective binary vector (PMR) is calculated; finally, the PMR is once again converted into a collective preference list, which can be transitive or intransitive. In the latter case we fall into a "Condorcet cycle". Since one of our goals is to escape from these cycles, we choose the initial condition ($t = 0$) in order to ensure that the initial collective list is intransitive.

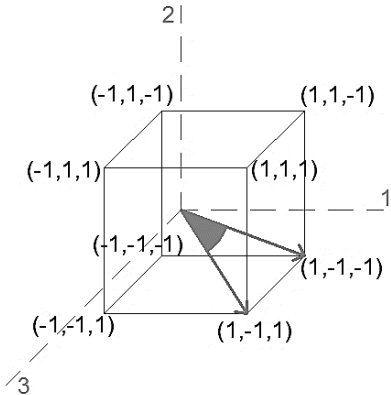

**Fig. 2.** Geometric interpretation of the overlap, as the scalar product between two binary vectors in the m-dimensional space (case of $n = 3$ alternatives, $m = 3$ components of the $8$ possible binary vectors).

However, as already said, we are interested in finding a final decision, represented by a collective list, which not only has to be transitive, but should also reflect quite appreciably the individual preferences of the stakeholders. This latter requirement can be verified by introducing a quantity called "overlap", that is a measure of similarity (or closeness) between any two lists. If $n$ is the number of alternatives and $m = n*(n-1)/2$ is the number of the possible pairwise couples, i.e. the number of components of each binary vector, the overlap is defined as:

$$O(i,j) = \frac{1}{m}\sum_{k=1}^{m} V_k(i) \cdot V_k(j) \quad (1)$$

where $V_k(i)$ and $V_k(j)$ are the $k$-th components of the two binary vectors $V(i)$ and $V(j)$ representing the preference lists of stakeholders $S(i)$ and $S(j)$. From this definition follows that the overlap $O(i,j)$ assumes values in the closed interval $[-1,1]$. In particular, if $V(i) = V(j)$, then $O(i,j) = 1$; if all the homologous components $V_k(i)$ and $V_k(j)$ have opposite signs, then $O(i,j) = -1$; if the components with opposite and equal signs are in the same quantity, then $O(i,j) = 0$. As an example, the similarity between two lists of $n = 3$ alternatives can be represented by the alignment between the two related binary vectors of $m = 3$ components in the m-dimensional space and the overlap coincides with the scalar product between these vectors; in this example, shown in Fig. 2, the two vectors are partially aligned and the overlap is 0.33.

The same formula can, of course, be used to calculate the overlap between any individual vector $V(i)$ and the *collective* vector $V(c)$, i.e. $O(i,c) = \frac{1}{m}\sum_{k=1}^{m} V_k(i) \cdot V_k(c)$. For example, it can be used at the beginning of a simulation (i.e. at $t = 0$) to evaluate the similarity between each individual list and the corresponding collective intransitive list, represented by $V^{t=0}(c_{intr})$; since the initial individual lists are randomly selected, one might expect that the average overlap of these lists with the collective list, calculated as $\bar{O}^{t=0}(i, c_{intr}) = \frac{1}{N}\sum_{i=1}^{N} O^{t=0}(i, c_{intr})$, assumes values close to zero.

*The max overlap strategy*
At this point, it is quite natural to imagine a first possible strategy to escape from the Condorcet cycle without any interaction among agents. Actually, at t=0, one could select, among all the possible transitive lists, the one whose binary vector $V^{t=0}(c_{tr}^*)$ has the maximum overlap with $V^{t=0}(c_{intr})$, i.e.:

$$O^{t=0}(c_{tr}^*, c_{intr}) = \frac{1}{m}\sum_{k=1}^{m} V_k^{t=0}(c_{tr}^*) \cdot V_k^{t=0}(c_{intr}) = \max\{O^{t=0}(c_{tr}, c_{intr})\} \quad (2)$$

Assuming that closeness between two cells in the collective preference space (see Fig. 1b) is a rough measure of their overlap, this strategy just corresponds to find, in this space, the "transitive island" (grey cell) closest to the intransitive initial list (bold white cell) (see Fig. 3a). Nevertheless, it is quite probable that the average overlap $\bar{O}^{t=0}(i, c_{tr}^*)$ between the selected transitive list and all the individual lists (proportional to the colour intensity of grey cells in the collective preference space) will be close to zero, like $\bar{O}^{t=0}(i, c_{intr})$. Actually, this is exactly what happens, as it will be shown later through the simulations. In other words, the final collective list found by the "max overlap strategy" is transitive, but it does not fulfil the requirement of appreciably reflecting the individual preferences (in terms of average overlap).

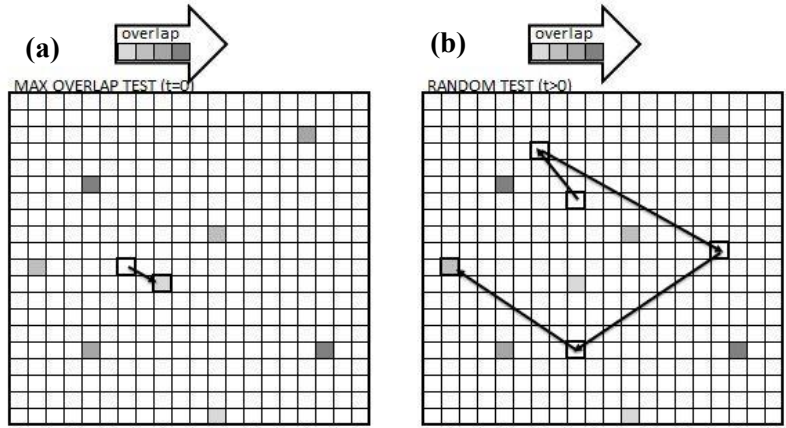

**Fig. 3**. Representation of (a) the max overlap strategy and (b) the random strategy in the collective preference space.

*The random strategy*
From the previous considerations it appears that, in order to get a transitive collective list with a higher average overlap with the individual lists, it is necessary a searching strategy able to find a darker grey island of transitivity in the collective preference space, starting from the initial cell of intransitivity. A random searching strategy has been tested to this aim. This implies the following dynamical algorithm: starting from the initial intransitive collective list, at each step $t > 0$ each stakeholder $S(i)$ randomly changes his preference list and the corresponding collective preference list is derived, as already explained for the previous strategy. The process ends when a transitive collective list is obtained, generally after a small number of steps. In the collective preference space, this corresponds to draw a "random path" in the sea of the intransitive lists, until a transitive island is found (see Fig. 3b). However, due to the randomness of this procedure, the corresponding average overlap $\bar{O}^{t=0}(i, c_{tr})$ results to be again very low. As simulations will confirm, even if the algorithm is further run - finding other randomly selected islands - there would not be an increase in the average overlap. This means that the random strategy fails too and that it is necessary to find a more suitable searching strategy, able to reflect the individual preferences.

*The opinion dynamics strategy*
As already said, interaction among stakeholders is a key of success for a transparent and shared decision-making process. In our simulations we try to mimic it through an opinion dynamics model that reproduces the opinion changing process in networks of connected and interacting people. In particular, at each step $t > 0$, each stakeholder $S(i)$ interacts only with his $N(i)$ first neighbours $\{S(j)\}_{j=1,\dots,N(i)}$ in the network. Due to the interaction, $S(i)$ has a certain probability of changing opinion, depending on both the influence $I(j)$ of his neighbours and the similarity of their lists. More precisely, we assume that $S(i)$ will change his opinion by adopting the list of his neighbour $S(k)$ with a probability $P(i, k) = \frac{I(k)}{\sum_{j=1}^{N(i)} I(j)}$, but only if the overlap $O(i, k) > 0$; otherwise, $S(i)$ will maintain his/her list. After all the stakeholders updated their lists at time $t$, a new PMR - and its corresponding collective preference list - is calculated: if the latter is transitive, the new average overlap $\bar{O}^t(i, c_{tr})$ is computed, recorded and plotted as function of time (Fig. 4 a); on the other hand, if during the process a Condorcet cycle occurs, the solution is discharged. In both the cases the algorithm goes on, in order to find more and more shared transitive solutions.

This procedure can be visualized again in the collective preference space, where an iterative and progressive path among transitivity islands, through adjacent intransitivity cells, is followed (Fig. 4 b). The striking point is that, contrary to what happens in the case of the random strategy, in general this path leads to transitive islands with higher overlap, until the "darkest" grey island is reached, i.e. the one with the maximum achievable overlap. In fact, as it clearly appears looking at the average overlap behaviour versus time (see again Fig. 4 a), the first transitive lists found usually show a low overlap, but, when the interaction is repeated, $\bar{O}^t(i, c_{tr})$ presents a growing – even if not monotone – trend, until it reaches a stationary state, corresponding to its maximum value (in agreement with the trajectory in Fig. 4 b). The final list is therefore assumed as the transitive "most shared" collective solution, appreciably reflecting the individual preferences of all the stakeholders.

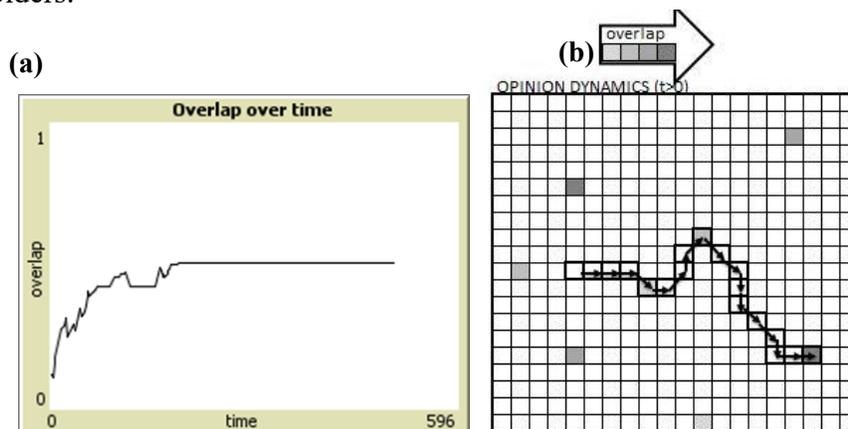

**Fig. 4** (a) Plot of average overlap over time and (b) representation of the opinion dynamics in the collective preference space.

The main routines described above are summarized in Fig. 5.

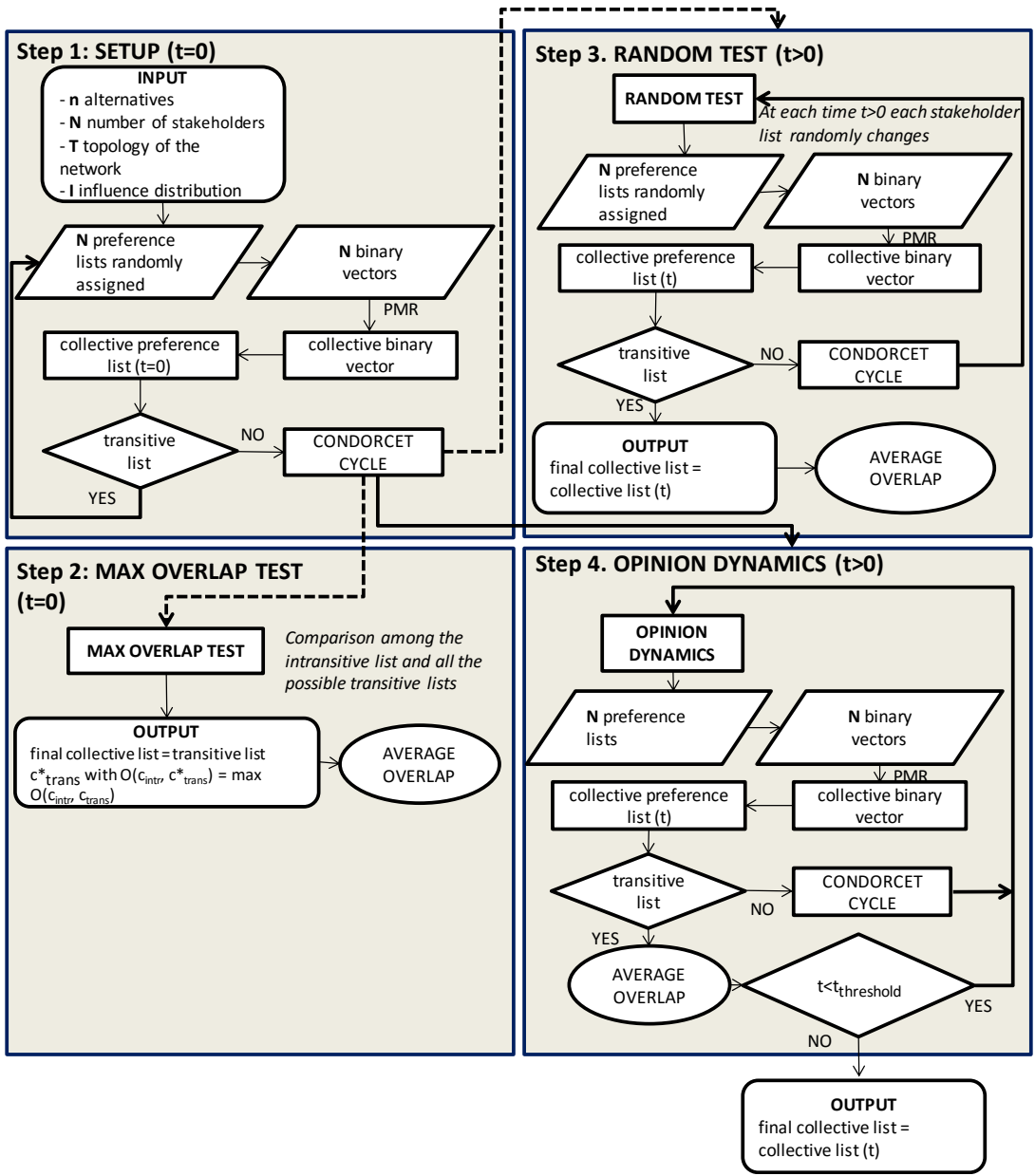

**Fig. 5.** Flow chart of the agent based algorithm.

### 3.2 Stakeholder networks description

Simulations aim at investigating how the collective decision-making process could be affected by the features of the participation process. In this respect, several simulations were performed by comparing the different strategies discussed in the previous section in order to select a transitive and shared collective decision, in particular when different numbers of stakeholders and different topologies are considered. Stakeholders are linked in social networks, i.e. graphs consisting of nodes (i.e. the social agents) and links (i.e. the relationships among them). Social networks are usually complex networks, whose structure can be of different kinds and dynamically evolving in time (Boccaletti et al., 2006).

Table 2 gives a short description of the different undirected networks used for the simulations (for a review see Estrada, 2011).

**Table 2** Networks used for the simulations. Node colours are different opinions at t=0, node radius is proportional to the node influence. The number of neighbours of a given node, i.e. the "degree" of the node, is indicated with k.

| Network | Description | NetLogo interface |
|---|---|---|
| Star | One central node (hub) is directly linked with all the other N-1 nodes. Every node has degree 1, except the hub that has N-1 | 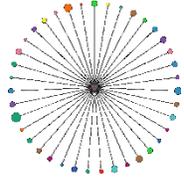 |
| Fully connected | It is a totally connected network where each node is connected with all the others. The degree is N-1 for all nodes. | 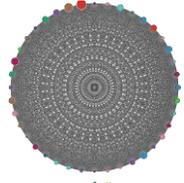 |
| Scale free Barabasi-Albert (BA) (Barabasi and Albert, 1999) | The network is built sequentially following the preferential attachment criterion: at each step a new node is added to an existing node with a probability proportional to its degree $k$. This process generates a network with a power law degree distribution, i.e. $P(k) \sim k^{-\gamma}$ with $\gamma = 3$. | 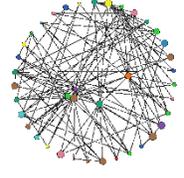 |
| Scale Free Tree with fully connected hubs | The network is built sequentially following a modified preferential attachment criterion where one path only exists for each pair of nodes, i.e. there are not "triangles of nodes" in the network. It can be therefore classified as a tree with some hubs, i.e. a few nodes with a very high degree. We further link all the hubs in a fully connected subgraph. This process generates a network with a power law degree distribution, i.e. $P(k) \sim k^{-\gamma}$ with $\gamma = 2 \div 3$. | 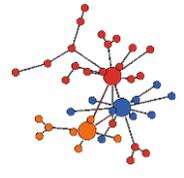 |
| Small world network (Watts and Strogatz, 1998) | Small world networks are "highly clustered, like regular lattices, yet they have short characteristic path lengths, like random graphs". Nodes are linked with their first 4 neighbours in a circle, with a certain probability $p$ of rewiring, i.e. to remove some links with the first neighbours and replace them with links pointing to random nodes (in general p = 2%). The average degree is 4. | 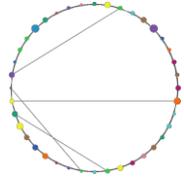 |

These prototypical networks are chosen to represent different stakeholder groups and the related interaction processes. The rationale behind the selected networks is that all these topologies can likely be found in real participation processes: (i) the star network is a typical structure of the Delphi practices, where a facilitator is linked with all the actors involved, but they are not linked to each other to avoid the risk of "leadership" (Dalkey and Helmer, 1963); (ii) the fully connected network can represent quite well the focus group meetings, aimed at analysing a specific topic with the interested stakeholders affected by the decision; (iii) the scale free BA shows a power law degree distribution with few high-connected nodes and many low-connected nodes and it is a typical structure found in many social networks (Barabasi and Albert 1999); (iv) the scale free tree with fully connected hubs also shows a power law degree distribution, but the high-connected nodes (i.e. the hubs) are all connected with each other and the subgroups form trees, representing typical hierarchical structures where the information from the nodes converges to the hubs, e.g.

those that can be found in citizen councils at different levels (neighbourhood, municipal or regional); (v) the small-world network is a structure that has been found in several real social networks, with high levels of communication efficiency thanks to the structure of regular network "rewired" with some long-range links (Watts and Strogatz. 1998). In addition, Bakht and El-Diraby (2014) demonstrated, considering infrastructure discussion networks (IDN) such as Twitter, that the structures of the interacting communities show characteristics of complex networks, in particular small-world behaviour, scale free degree distribution, and high clustering. This type of engagement with social media methods is also known as "Microparticipation" (Evans-Cowley and Griffin, 2003) and can be very interesting for the study and interpretation of the complex phenomenon of social interaction.

In all the networks considered for our simulations, the influence $I(i)$ of node $i$-th is an integer random variable distributed with a Poisson law with mean 8. The simulations performed considered $n = 6$ alternatives for each preference list, while various sets of simulations run in correspondence of different number of nodes ranges (from 20 to 120) and of different network topologies. In general, several simulation runs with the same structure but different initial conditions were considered for each case. In particular, the results of each simulation were averaged over 100 events, each of them running over 500 time steps, that have been verified to be enough for reaching the overlap stationary state shown in Fig. 4a.

### 3.3 Simulation results

Results are expressed in terms of two order parameters: (i) the asymptotic value of the average overlap $\bar{O}^t(i, c_{tr})$, i.e. a measure of the similarity among the individual lists and the final collective one after interaction, and (ii) a parameter representing the "interaction efficiency" ($IE$), calculated dividing the asymptotic average overlap by the number of links $n_l$, in order to include the "communication costs" to be sustained in highly connected networks:

$$IE = \frac{\bar{O}^t(i, c_{tr})}{n_l} \qquad (3)$$

Both these measures were also averaged over 100 simulation runs.

In Fig. a, the (averaged) asymptotic average overlap is plotted as function of the network size $N$ and for all the previously introduced network topologies. All the scores range from about 0.3 to 0.6, that is a very good result. In particular, they stay quite high ($> 0.5$) when the size of the networks is small ($N = 20$), regardless of the topology. When the number of stakeholders increases some differences among the different networks do emerge. As expected, the fully connected network performs very well, because each stakeholder influences, and it is influenced, by all the others in the network. The same holds for the star topology, probably because there are, on average, only two degrees of separation between any couple of nodes. On the other hand, the small-world network shows the lowest values of final overlap for $N > 50$, and – in the same range – also the performance of the scale free tree with fully connected hubs is not excellent, likely due to the absence of interaction triangles among stakeholders; at variance the overlap outcome for the scale free BA network remains quite constant at high levels. These results represent a first confirmation of the robustness of the interaction strategy in finding the more shared transitivity solutions in the collective preference space. But it is also interesting to see what is the behaviour of the interaction efficiency IE for the same networks, obtained dividing the asymptotic overlap by the number of links for each topology.

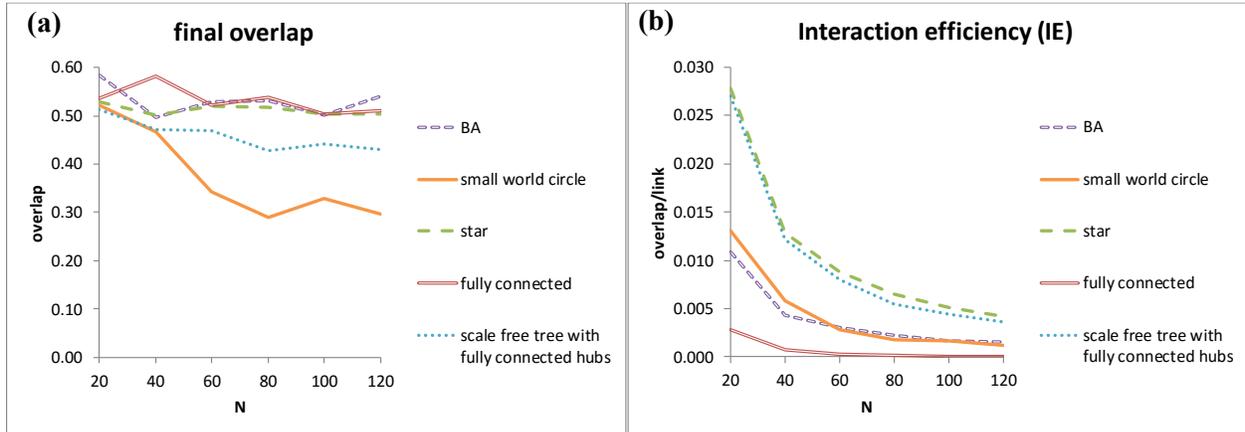

**Fig. 6.** Plots of (a) the final overlap and of (b) the normalized final overlap, called "interaction efficiency", resulting from the interaction dynamics performed for different topologies and for an increasing number of stakeholders.

As shown in Fig. 6b, if one includes the "communication costs" in the balance, for small groups of stakeholders ($N = 20$) the fully connected network becomes the less efficient if compared with the other topologies, while the star and the scale free tree network show the highest values of overlap per link. In any case, increasing the number of nodes, differences among the topologies diminish very quickly. Looking more in detail, in Fig. the asymptotic overlap values for each one of the 100 simulation runs are plotted for the various topologies and for the two limiting sizes of the networks (120 nodes, top panels, and 20 nodes, bottom panels), in order to appreciate their variability: in all the cases considered, the overlaps appear to strongly fluctuate around the average values already reported in Fig.6a (here indicated by an horizontal full line), with standard deviations ranging from about $\pm 0.18$ for larger networks to about $\pm 0.21$ for smaller networks. However, the asymptotic overlaps stay always quite higher than those obtained with the other two strategies, i.e. the "random" test and the "max overlap" test, also reported in the bottom part of all the panels.

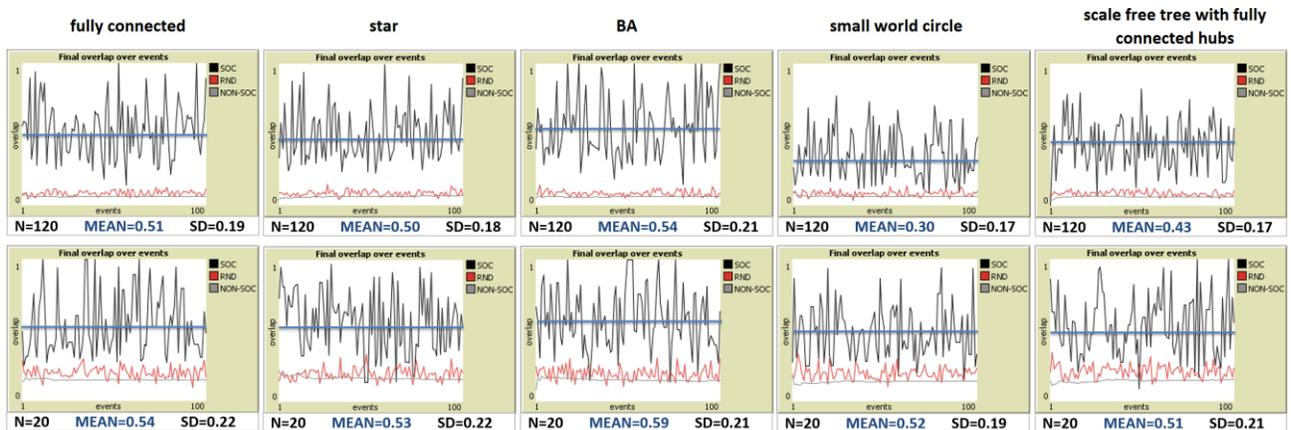

**Fig. 7.** Plots of the asymptotic overlap over events for all the network topologies with 120 nodes (top) and 20 nodes (bottom) with the three strategies ("SOC" = socialization or interaction strategy; "RND" = random strategy; "NON-SOC" = max overlap strategy).

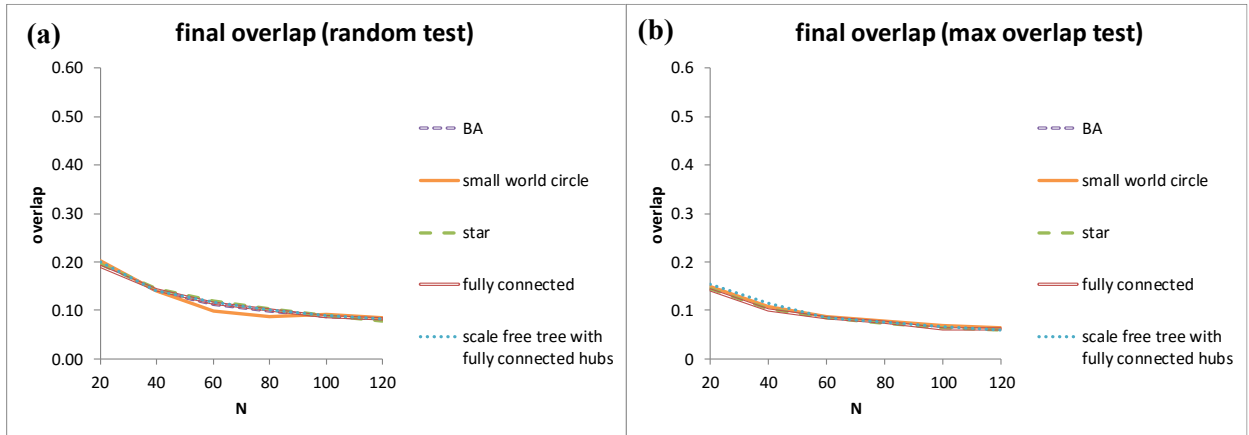

**Fig. 8.** Plots of the final overlap (averaged over 100 events) resulting from the application of the random strategy (a) and the "max overlap" strategy (b) for different topologies and an increasing number of stakeholders.

The latter result is confirmed by the plots in Fig. 8, where the averaged values of the asymptotic overlap are reported for both the "random" strategy (a) and the "max overlap" strategy (b), again as function of the number of stakeholders and for the various topologies considered. In both cases the final overlap is always quite lower than the one resulting from the interaction process and it is also independent of the topology (since both the tests do not involve any communication among the agents). In particular, with the first strategy the overlap values range from 0.08 to 0.23 while, with the second strategy, they go from 0.06 to 0.15.

Summarizing, these findings confirm that, even if it is possible to imagine other strategies that allow to circumvent the "Condorcet paradox" faster and easier than the interaction one, only the collective solutions that come out from repeated discussions and opinions' exchanges among the stakeholders are able to actually reflect enough the individual preferences, ensuring the "most shared" transitive decision, whatever the topology one decides to adopt for the interaction network. On the other hand, taking into account the topology of interactions, the previous results could also give some insights or suggestions in order to build an efficient participation process. Actually, they can be useful from a practical point of view because the real structures of many participation processes share similarities with the analysed interaction networks. Preliminary proofs derive from some participation experiments that were carried out for a first attempt of model validation as explained in the following subsection.

### 3.4 Towards a validation of the model

The problem of ABM validation is controversial and it is still a challenging task (Windrum et al., 2007; Moss, 2008). A good data basis is crucial for a good model (Buchmann et al, 2016), but it can also be useful to directly involve stakeholders in the building of the model, i.e. the "companion modelling" approach (Bousquet et al., 1999; Voinov and Bousquet, 2010).

For a first step towards validation, the results of *ad hoc* participation experiments are used to test the suitability of the model in reproducing participatory decision-making processes (Le Pira et al., 2015b, 2015c). A first case study in our case was a participation experiment with students at the University of Catania (Italy), where a two-round interactive AHP was set up to elicit their preferences about mobility management strategies to be adopted in the university campus and to see how interaction among them could change their opinions; results of one round of real interaction among a "quite homogeneous community of stakeholders" are in agreement with those obtained with a simulation that reproduced a repeated interaction starting from random individual preferences (i.e. "heterogeneous community") (Le Pira et al., 2015b). In particular, the values of the final average overlap were almost the same (0.57 from our simulation versus 0.59 from the experiment), while the "Condorcet paradox" did not show up in the real experiment thanks to the opinions of the group that were quite homogeneous. The simulated network was a fully connected one with exactly the same number of nodes used to mimic the interaction process that involved all the students in a plenary session.

The second case study was a combined Delphi-AHP experiment about cycling mobility in Catania (Italy) where a panel of experts from the University and stakeholders (from the municipal transport company and cyclists' associations) analysed and expressed opinions about possible strategies to increase the use of bicycles. The Delphi structure allowed an anonymous interaction among the actors and led to a convergence of opinions. The simulated network was a "star network", where the hub (i.e. the facilitator) was directly linked with all the others. This could represent quite well the anonymous interaction process among the members of the panel, that could not communicate with each other but knew the other (averaged) answers. In this case the model was fed with data about the actual initial opinions of the stakeholders and the opinion dynamics was properly modified to mimic the real opinion changing process. Actually, the opinion dynamics process is simplified with respect to reality, since a collective preference ranking is evaluated through PMR (based on stakeholders' real preference rankings) and then agents decide to align their ranking to the collective one according to the similarity with it (in terms of overlap); on the contrary, in the real experiment stakeholders were asked to "adjust" their preferences if they fell out of a range represented by the majority of them. Despite its simplicity, results of the simulation were in agreement with reality (e.g., final overlap value obtained in the simulation was 0.88 versus the value 0.80 obtained in the experiment) and show that the model is able to suggest how to guide a Delphi method to reach a good convergence of opinions and avoid the risk of "Condorcet paradox" (Le Pira et al., 2015c).

In conclusion, we are fully aware that a real participation interaction is not easy to control and sometimes it is quite difficult to draw out constructive conclusions from it. Nevertheless, the results of the simulations are not intended as a tool to build real networks or to "guide" the individual opinions, but they demonstrate to what extent the degree of interaction and the type of communication have an influence on the final collective decision. In our opinion, the optimum for a participatory planning approach can be reached neither with dictatorial single-judge processes, nor with diffuse participatory democracy, but with a mixed interactive process where the topological structure of the interacting stakeholder network can be guided by the simulation results.

## 4. CONCLUSIONS

Complex public decisions with widespread collective impacts affect multiple stakeholders and require a participatory approach to decision-making.

In this paper we have shown, through an ABM which implements an opinion dynamics on networks with different topology, the effectiveness of a repeated interaction among stakeholders in order to find a collective decision, which at the same time is consistent and shared. The opinions are expressed in terms of preference lists of alternatives and a collective list is derived from aggregation of the individual ones avoiding intransitive solutions, i.e. the "Condorcet paradox". The preference lists are initially randomly assigned to stakeholders, being the only way to guarantee a shared decision not being influenced by the existence of strong positions towards certain directions. Simulation results show that, even if there are other possible strategies to avoid decision deadlocks caused by the "Condorcet paradox", only a repeated interaction among stakeholders allows to reach a good degree of convergence of opinions, numerically assessed in terms of overlap among the lists. Most of the social networks topologies considered guarantee good communication flows and help interaction towards a shared decision; nevertheless, the degree of connection largely affects the "communication costs", with relevant differences in the interaction efficiency. Results of the simulations are also in agreement with some participation experiences regarding transport decisions, representing a first attempt to validate the model. Further research will tend to improve the reliability of the model with empirical data and its applicability to transport decision-making contexts.

The model could be of help for policy-makers and practitioners to support stakeholder-driven decision-making processes about complex decisions, and could be successfully adopted also in other contexts different from transport planning in order to reach within a reasonable time a more participated decision.

**ACKNOWLEDGEMENTS**
This work was partially supported by the FIR Research Project 2014 N.ABDD94 of Catania University.